\documentclass[12pt,fleqn,oneside]{article}

  \newtheorem{theorem}{Theorem}
 
\newtheorem{nn}[theorem]{$\!$\hspace{-2pt}}
\newtheorem{apnn}[theorem]{{\bf A\hspace{-2pt}}}

  \newenvironment{proof}{{\it Proof.\ \/}}{\  }
  
  \newcommand{\bb}{\begin{nn}\em }
\newcommand{\ee}{\end{nn}}

    \newcommand{\be}{\begin{enumerate} }
\newcommand{\een}{\end{enumerate}}
  
   \newcommand{\apbb}{\begin{apnn}\em }
\newcommand{\apee}{\end{apnn}}
\renewcommand{\a}{\`{a}\ }

 \newcommand{\up}{\uparrow}

      \newcommand{\Ra}{\Rightarrow}

  \newcommand{\n}{~}
  
 \newcommand{\ap}{'}

 \newcommand{\ra}{\rightarrow}
  
 \newcommand{\db}[1]{ [\![{#1}]\!]}
 \newcommand{\dpa}[1]{[{#1}]}

\newcommand{\ld}{\mbox{{\footnotesize \ldots}}}

 \newcommand{\noind}{\!\!\!\!\!\!\!\!\!\!\!\!\!\!}

\newcommand{\TM}{{\footnotesize TM}\ }
\newcommand{\tms}{{\footnotesize TM}}

\newcommand{\ang}[1]{\langle{{\tt #1}}\rangle}
\newcommand{\IH}{_{\,\!_{{\rm I.H.}}}}

  \newcommand{\FN}[1]{\mbox{{\footnotesize{\rm  {#1}}}}}

\begin{document}
%%%%%%%%%%%%%

 \thispagestyle{empty}
 \pagestyle{plain}
 \title{{\Large
 A  Predicative % recursion scheme
  Harmonization \\ of  the Time and Provable Hierarchies}}
  
   \author{
    Salvatore Caporaso\\
     {\small Dipartimento	di Informatica dell\ap Universit\a di Bari}\\
   {\small  caporaso@di.uniba.it }
    }

    \date{}
 
  \maketitle

{\small
{\bf Abstract}
\quad
	A decidable transfinite hierarchy is defined by 
	assigning ordinals to the programs
	of an imperative language. 
		It singles out: the   classes $\FN{TIMEF}(n^c)$
			and 	$\FN{TIMEF}(n_c)$;  
the finite	Grzegorczyk classes
at and above the elementary level,
	and the $\Sigma_k$-IND fragments of PA. 
	Limited operators, diagonalization, and majorization
	functions are not used. 
}

\section{Introduction}
\thispagestyle{empty}  

\bb \label{TH}
 {\bf Motivation}\quad
Most  transfinite hierarchies ${\cal Z}_\alpha$ are   defined in two 
steps: a sequence $Z_\alpha$ 
 	    of   majorization functions
 	    is introduced by means of a recursive operator 
 	    at successors, and,  at limits $\lambda$, by
 	    a {\it diagonalization\/} of the form $Z_\lambda(n)=Z_{\lambda[n]}(n)$;
	  	the  class of functions ${\cal Z}_\alpha$ is then obtained by
	  	closure  of $Z_\alpha+\bigcup_{\beta<\alpha}{\cal Z}_\beta$
	   under limited \FN{PR}
% 	   \n\cite{Rose}
	   or product,
% 	   \n\cite{FW},
	   and substitution (often limited too).
The
 rate of growth of such ${\cal Z}_\alpha$ 	is either slow to the point of dispersing
 a single  class 
	like  $\FN{TIMEF}(n_k)$ along an $\omega_k$-type segment;
	or fast to the  point of ignoring the  complexity classes.

	The  Implicit Computational Complexity (ICC) 
	program is looking  for  characterizations: (a)
	 obtained by closure  under 
	 resource-free (that is, {\em unlimited\/}) 
	 schemes
	  of a small  stock of basic operators, not including {\em ad 
	  hoc\/} functions;
	   (b) whose
	     membership  should be decidable syntactically. 	    
	    For example, Cobham\ap s 
	     polytime 
	    disagrees with (a)  because of
	    the {\em smash\/} function; and  with (b) because  
	     limited \FN{PR} on notations is undecidable.
	   The same applies to the role of  $Z_\alpha$ and
	       of the limited schemes in the transfinite hierarchies.

% 	 promoted and developed by Leivant and others after the 
% 	 results \cite{Lpoly,BC},	
% 		many classes ${\cal C}$ have been revisited   
% 		in terms of imperative languages (see  the proceedings of the yearly	    
% 	 ICC workshops; see also\n\cite{B3,CZG}).
% 		We all shared in common two main issues: 

	        ICC was also dubbed  {\em predicative\/},
	       on the grounds of the analogy,  pointed out by Leivant, between
	       growth of sets and impredicative comprehension, on one hand; and 
	       growth of functions and nested recursion,
	       on the other. %Subrecursive hierarchies 
% 	       should be built-up by {\em stages\/}, in the sense that,
 	       Like in other branches of mathematics,
 	        impredicative recursion 
 	        can be reduced by means of constructions by {\em stages\/}: at $\alpha$,   only  
	         functions introduced at  stages  $\beta<\alpha$ can be used.
	       How to do this is less easy at limits $\lambda$ than at successors.  
A first solution is using {\em enumerators\/} \
	       $e\in{\cal Z}_{<\lambda}$, 
	         to put  $f\in{\cal Z}_\lambda$ for any  $f$ such that  (Rogers notation) 
	       \[
	       f(n)=\varphi_{e(n)}(n)\quad\mbox{provided 
	       that}\quad\varphi_{e(n)}\in{\cal Z}_{<\lambda}. 
	       \]
	        Membership is then undecidable,
	       because of the condition about $\varphi_{e(n)}$. Moreover, a
	        natural way    to compute   $Z_\lambda$ is via the form
	       $\varphi_l=\varphi_{e(l)}$  of the recursion theorem.
	       But  then  a function  {\em apply\/} is needed,
	       and it is not clear to which stage does it belong. 
	     
	    In last decade we have been  trying to collect and 
	    connect, in a same taxonomy or hierarchy,
	    built-up by means of a same predicative criterion,
	       as many complexity and subrecursive 
	    classes as possible. In our contributions to four ICC workshops, 
% 	    \cite{C2} 
  and	    elsewhere \cite{C,CZG}, a number of hierarchies were 
 introduced, covering the PR, the
	    elementary, and the exponential time classes.
	    These exercises were based on constructive forms of diagonalization,
	    based on linear time enumerators.  
	    Apply-like functions were not used, but membership was undecidable in 
	    these cases too.

	\ee
\bb{\bf Statement of the result}\quad  
	  In this paper an imperative language
	  is defined  by closure of a generic stock of constant time programs
under  composition and under a new {\em repetition\/} scheme 
(see\n\S\ref{comment})¥.
	  A
	    hierarchy ${\cal C}_\alpha$ 
($\alpha<\varepsilon_0$) is defined 
by assigning tree ordinals (see \cite{FW}) to its programs.
It singles out  at $\omega^c$ and $\omega_c$ the   classes
$\FN{TIMEF}(n^c)$ and $\FN{TIMEF}(n_c)$, and, at higher ordinals, 
the functions provably total in systems like PA.
Besides harmonizing  {\it small\/} and {\it big\/} classes, this hierarchy is
predicative, decidable,
 by closure under unlimited operators,
 and it doesn\ap t use any form of diagonalization. 	       
 The language is built-up without making use
 of   majorization functions;
  the Wainer-Schwichtenberg functions enter the scene later,
  to play the role of uniform scales against which
 we can measure the power of its decidable fragments. 
 \ee
\bb\label{th}
	{\bf Theorem}\quad We have 	
	($\FN{TIMEF}(f(n))\approx{\cal C}_\alpha$ 
	means that 
 ${\cal C}_\alpha$ {\it sandwiches\/} between 
 $\FN{TIMEF}(f(n-1))$ and $\FN{TIMEF}(f(n+3))$)
% 	We have %the following characterizations: 
% 	($c\geq 1$, $d\geq 0$, $\omega^2\leq\alpha<\varepsilon_0$)
		\[\noind\!\!
		\begin{array}{lrlllllllll}
	 \mbox{Polytime}&\FN{TIMEF}(n^c)&=& {\cal C}_{\omega^c}&(c\geq 1)  
	 \\
	 \mbox{Superexponential time}\qquad
	&\FN{TIMEF}(n_c)&\approx& {\cal C}_{\omega_{c}}
	\\
	 \mbox{Grzegorczyk  classes}& 
	 	\FN{TIMEF}(F_{c+2}(n))&\approx&{\cal C}_{\omega^{\omega+c}}
\\
%  	 (c\geq 1).
 \mbox{2-nested recursion}& 
\FN{TIMEF}(F_{\omega c+d}(n))&\approx& {\cal C}_{\omega^{\omega(c+1)+d}}
	&(d\geq 0)
	 	\\
	\mbox{Higher classes}&\FN{TIMEF}(F_{\alpha}(n))&
	\approx& {\cal C}_{\omega^\alpha}&(\omega^2\leq\alpha<\varepsilon_0).
	\end{array}
\]
% where
%  for  $e=0$ if $\alpha<{\omega^2}$, and $e=1$ otherwise. 
% %  $\alpha\geq{\omega^2}$.
%    
See \ref{thehierarchy} for a more precise definition of ${\cal C}_\alpha$
and $\FN{TIMEF}(f(n))$; see  \ref{WS} 
for the Wainer-Schwichten\-berg functions $F_\alpha$; and 
see\n\ref{proof-th} for the proof of this theorem.
 	 \ee

 \bb{\bf Related work}\quad \label{}
 A harmonization of the  computational complexity hierarchies
  with the subrecursive ones, in terms of
imperative programs, appears to be lacking.
Moreover, we are not aware of other imperative languages 
 for the whole class of the provable functions which 
 replace diagonalization by  a {\it 
 finite\/} number of  schemes. 
 In proof-theoretical terms, we have\n\cite{LeivantT} and the  extension
		 of\n\cite{FW} to the  
	small classes contained in\n\cite{OW}.

Polynomial time is associated with programs whose repetion nesting depth is 1.
The scopes of their  repetitions are  iterated
for a number of times equal to the input length.
In this way the distinction
between nested and un-nested repetitions  mirrors the
 distinction \cite{BC} between safe  and ordinary variables.

 The hierarchy  can single-out   the   super-exponential classes
 because the  repetition scheme is quite {\it honest\/}:
 runtime for the simulation by \TM  of $f(n)$ is $f(n+3)$  
	  	   (in literature one finds   $2^{f(f(n))}$ in\n\cite{FW}
 and  $2^{f(n+|\alpha|)}$ in\n\cite{W}). 
\ee

\section{The Language}

  \begin{nn}\label{synt}\em
     	 {\bf Syntax}\quad
     Let a class {\bf D} of  data be given,
     together with an appropriate  measure $|x|$ of the {\it length\/}
     of all $x\in{\bf D}$. 
     	   Let {\bf A} denote a generic, finite collection  of {\em initial\/}
     		programs,
     		defined on 
     		 {\bf D}.  To avoid tedious analyses of marginal cases
     		 we assume

     	(a)\quad
     	$|{\tt A}(x)|=|x|+1$  for  each  ${\tt A}\in{\bf A}$ and    $x\in{\bf D}$;
      	
     	(b)\quad each ${\tt A}\in{\bf A}$
     	 is simulated in a constant time
     	by a \tms; 
     	
    (c) \quad       	 $|x|\geq 2$ for all $x$. 
     	 \label{+1} 
  \label{B-A}  	
  
Define a class  ${\bf A}^*$ of programs  by 
  closure 
 of ${\bf A}$ under
  the  {\it composition\/}  and   {\it repetition\/} schemes
 \[       	  {\tt Q\ R}  \qquad\qquad
      	 \langle{\tt  Q}\rangle\qquad\qquad\qquad\qquad({\tt Q,R}\in{\bf A}^*).
      	 \]      	    
\label{acyclic} 
A program is
  {\em acyclic\/} if it is a composition  of initial programs;
   {\tt c} will denote a composition of  $c$ initial programs
    (so we always have ${\tt 1}\in{\bf A}$).
    A program is   {\it safe\/} if its repetitions (if any) don\ap t occur in the scope of other 
 repetitions. 
% Its form is then (by abuse,  $\ang{0}$ is {\tt 1})
% \begin{equation}\label{safe-form}
% 	\ang{d_1}\ldots\ang{d_k}\qquad\qquad(k\geq 1;\ d_i\geq 0;\ 1\leq i\leq k)
% \end{equation}
% 
\end{nn}
  \bb     \label{notat-ttn}
    {\bf Notation} 
 \quad     	     	   
   $i,\ldots,q$ are numerical variables,
    	while    $c,\ldots,e$ are numerical para\-meters.
    	{\tt P},\ldots,{\tt T} are programs, and  {\tt A}, {\tt B},\ldots are 
    	initial programs.
    	$\epsilon$ is the empty string over the current alphabet, or it is the absent program.
    	${\tt P}^0$ is  $\epsilon$,
   and  ${\tt P}^{c+1}$ is ${\tt P}^c\;{\tt P}$.

     $\alpha,\beta,\ldots$, and  $\lambda,\mu$ are
(resp.) tree-ordinals, and limits.
 $\alpha_0\dpa{\beta}$ is $\beta$,
 $\alpha_{c+1}\dpa{\beta}$ is $\alpha^{\alpha_c\dpa{\beta}}$, and
$\alpha_c$ is $\alpha_c\dpa{\alpha}$. So $\alpha_2\dpa{3}$ is 
$\alpha^{\alpha^3}$, and $\alpha_1$ is $\alpha^\alpha$.

Throughout this paper,   $n$ is the length of the current $x$, and $l\geq 3$ is 
$n+1$.
\label{dlo}
\ee
 \bb
%  {\bf Definition}
\ (1) \
The {\em length\/}  $|{\tt P}|$ of {\tt P}  is the overall number of its
initial programs and repetitions.

 (2)\quad The  {\em depth\/} $\Delta\,{\tt P}$ of {\tt P} is the nesting depth 
 of its   repetitions.

 (3)\quad  Define the  {\em tree ordinal\/} (see \cite{FW})  $o\,{\tt P}$ of {\tt P}
 by  
 \[
 o\,\epsilon=0;\qquad
 o\,{\tt 1}=1;\qquad o\,{\tt Q  \,R}= o\,{\tt R}+ o\,{\tt Q};
\qquad o\,\ang{{\tt Q}}=\omega^{o\,{\tt Q}}.
\]
So 
$o\,\ang{1}\ang{2}=\omega^2+\omega$;\quad
 $o\,{\tt 1}\,\ang{\ang{2}}=\omega^{\omega^2}+1$; 
  and $o\,\ang{1\ang{\ang{1}}1}=\omega^{1+\omega^\omega+1}$.
% $o\,{\tt 1}\ang{3}\ang{2}{\tt 4}=4+\omega^2+\omega^3+1$.

For all tree-ordinals $\alpha\neq 0$ and $\beta\neq 0$ we have

$
 {\rm (a)}\quad   
 \alpha\mbox{ is in Cantor normal form 
iff $\alpha<\varepsilon_0$};
\quad 
{\rm (b)}\quad 
\alpha,\beta<\alpha+\beta\mbox{ and }\alpha<\omega^\alpha. 
$

By  (a) a   program of the form {\tt P\,Q} with
{\tt P} {\em stronger\/} than {\tt Q} (in the sense that
$o\,{\tt P}>o\,{\tt Q}$) does not interfere with our hierarchy. By  (b)
 an induction on $o\,{\tt P}$
is tantamount to an induction on the construction of {\tt P}.

 {\bf Notation}\quad 
 $|o\,{\tt P}|=|{\tt P}|$ (that is $|\alpha|=|{\tt P}|$ iff $\alpha=o\,{\tt 
 P}$). 
% This is a 
% a slight variant of the measure known as {\it Cichon norm\/} 
% \cite{cichon}.
\ee

\bb
\label{LIF} 
Any program {\tt P} is either in the form ${\tt A\,S}$ or  in the form
 \begin{equation}
 \label{NP-long}
\ang{\ang{\ldots\ang{\ang{1Q}R_1}\ldots}R_c}\ {\tt S}\qquad\qquad (c\geq 0;
	\mbox{ {\tt Q, S}, and any ${\tt R_i}$ may be }\epsilon).
	\end{equation}
	Hence we may define the {\em normal parsing}   of any {\tt P}
	(not acyclic) by
% 	 such that % 	$o\,{\tt P}\geq\omega$ 
\[
{\tt P= d}\ \langle^{c} \langle{\tt 1\;Q\rangle\ U}
\quad\quad\qquad\qquad(d,c\geq 0;\mbox{ {\tt Q} and {\tt U} possibly 
absent})
	\]
where {\tt U} is a string over ${\bf A}\cup\{\ang{,}\}$
which is not necessarily a well formed program.

For example, in the normal parsing of $\ang{\ang{{\tt 2}}}$
we have $d=0$, $c=1$, ${\tt Q=1}$ and ${\tt U}=\rangle$.
%   where  $\langle^{c}{\tt U}$ is a 
%    well-formed program. Note that, for $c=0$,  {\tt U} is well-formed or absent.
\ee

      	    \begin{nn}\label{semantics}
     	{\em {\bf Semantics}\quad An {\em instantaneous description\/} (\FN{ID}) 
is an expression $D$ of the form
${\tt P}:z$. 
     	Next       	    	 rules
      	    	take any \FN{ID}       	    	
into the next one 	 ($c\geq 0$; {\tt  A, Q} and {\tt T} are not absent)
% , while {\tt 
% A, Q} and {\tt T} may not) 
      	    	\[%\noind
      	    	\begin{array}{rlcrlcrcrlllrlrrllrllr}
      	    	 \langle^c	\langle {\tt A\,Q  \rangle \  U} &:x&\ra &
      	     \langle^c 	\langle {\tt Q  }\rangle ^{l}\  {\tt 
      	     U}&:x&&
      	    		\mbox{{\it R-eduction\/}}\\
      	    			 \langle^c	\langle {\tt A  \rangle \  U} &:x&\ra &
      	     \langle^c {\tt A}^{l}\  {\tt 
      	     U}&:x&&
      	    		\omega\mbox{{\it -elimination\/}}\\
 {\tt A  \,R}&:x&\ \ra\ &{\tt R}&:{\tt A}(x)\quad&\ (\Delta\,{\tt R}\neq 1)\ &
       	    		\mbox{{\it A-pplication\/}}\\
 {\tt A  \,T}&:x&\ \ra\ &{\tt T\, A}&:x&(\Delta\,{\tt T}= 1)&
       	    		\mbox{{\it safe P-ostponement\/}}\\
       	    		       	    		\end{array}
       	    		\]  
      	    		The {\bf computation\/} ${\bf C}({\tt P},x)$
	of ${\tt P}$ on $x$ is the sequence
	\[
	D_0={\tt P}:x,\;\ldots,\;\epsilon:y=D_p\qquad\qquad\qquad
	(D_i\ra D_{i+1}\quad
      	    	\mbox{for all}\quad i<p).
	\]
      	    
       	    {\bf Notation} \ (1) \
      	     	 $D\Ra D^*$
      	    means that there is  a  {\em (sub)computation\/} of the form
      	           $D,\ldots,D^*$.   	    	 
      	    		${\tt P}(x)=y$  stands for
      	    		 ${\tt P}:x\Ra\epsilon:y$.
      	    		 
      	    		 (2)\quad For any numerical function $f(n)$,
      	    		   ${\tt P}(x)=_cf(n)$ is short for
      	    		    $|{\tt P}(x)|=n+f(n)$.

      	    		 So, we may now write  clause \ref{+1}(a)
      	    		 in the form ${\tt A}(x)=_c1$
      	    		 (or, by abuse, ${\tt A}:x=_c1$). 
  }
  \end{nn}

\bb{\bf Comment}\label{comment}\quad
The {\it leftmost\/} repetition  $\ang{A\,Q}$ of any program {\tt P} is processed 
first,
and replaced by $l$ copies of a less complex program, namely by:
$\ang{Q}^l$
if {\tt Q} is not absent;  and ${\tt A}^l$, if ${\tt Q}=\epsilon$.
Assume that, after unnesting 
of some repetitions, {\tt P} has been reduced to the form $\ang{A}\,{\tt R}$.
By rule $\omega$ we obtain ${\tt A^{\it l}\,R}$. 
At this point, if   {\tt R} is not safe, rule $A$ is applied $l$
times and the  length of the  intermediate result is doubled.
On the other hand, if {\tt R} is safe, rule $P$ delays any change to  $x$ until all
repetitions have been  unnested.
%  and reduced to an acyclic program.
  In this way we
 keep $x$  {\it safe\/}
in the sense of  \cite{BC}, and we ensure that all  iterations
are repeated for a  number of times that equals the input length. 

If  we have ${\tt P=S\,T}$ then {\tt S} is processed  until it vanishes
(if {\tt T} is not safe) or until it has been reduced to a postponed acyclic program. So we have 
\[
{\tt S\,T}:x\Ra {\tt T:S}(x)\quad\mbox{if {\tt T} is not safe;\quad}
{\tt S\,T}:x\Ra {\tt T\,p}:x\mbox{ and }{\tt S}(x)=_cp\quad
\mbox{if {\tt T} is}.
\]
 \ee	  	    		      	    		
\begin{nn}\label{Ex1}
	{\em 
  	{\bf Example}\quad
%   	Put ${\bf D}=\{1^l|l\geq 1\}$, and  ${\bf A}=\{{\tt S}\}$
%   	where  ${\tt S}:x=x\,1$. 
%   	Writing $\ang{{\tt c}}$ for $\ang{{\tt 
%   	S}^c}$,
We have  
\[\noind\!\!
\begin{array}{lrllr}
1&	\ang{{\tt 1}}:x&\Ra&{\tt l}:x&\mbox{(rule)  $\omega$ with }{\tt 
U}=\epsilon\mbox{ and }c=0\\&&=_c& l&
 A\mbox{ ($l$
	times) since }\Delta\,{\tt i}=0\mbox{ for all }i\\
	 2&	\ang{{\tt 1}}^{d}:x &\Ra&
 	 {\tt l} \ang{{\tt 1}}^{d-1}:x
%  	 {\tt n\ap}\ \ang{{\tt 1}}^d:x\quad\quad
 	 &\omega\mbox{ with }{\tt U= \ang{{\tt 1}}^{d-1}}
   	\\&&\Ra&
 	 \ang{{\tt 1}}^{d-1}\ {\tt l}:x
%   	 {\tt n\ap}\ \ang{{\tt 1}}^d:x\quad\quad
 	 &P\mbox{ ($l$
	times) since }\Delta\,\ang{{\tt 1}}^{d-1}=1
 	\\&&\Ra
 	 &  \ang{{\tt 1}}^{d-2}{\tt l}^2:x
 	 &\omega\mbox{ and }P\mbox{  since }\Delta\,\ang{{\tt 1}}^{d-2}\,{\tt l}=1
 	 \\&&\Ra&
 	 \ldots\ \Ra \ {\tt l}^d:x\qquad\qquad 
 	  	\\&& =_c&dl
 	  	&\mbox{$A$ for $dl$ times since $\Delta {\tt i}=0$ for all $i$}\\
 	 3&	\ang{2}:x&\Ra&
     \ang{{\tt 1}}^{l}:x 
 	 &R\mbox{ with }{\tt Q=1} \mbox{ and }{\tt U}=\epsilon
 	 \\
 	 &	& =_c  &l^2
 	 &\mbox{part 2 with $d=l$.}
\\
4&	\ang{\ang{{\tt 1}}}:x&\Ra&
 	 \ang{{\tt l}}:x\quad\qquad\qquad&\mbox{$\omega$, with  $c=1$, ${\tt Q=U}=\epsilon$}\\
 	& &=_c &  	  	 	l^{l}&\mbox{because the}
 	 	\end{array}
\]
 forthcoming  Lemma\n\ref{poltime} 
will generalize part 3, by proving
$\ang{{\tt d}}(x)=_c l^d$. 
}
\end{nn}

    \bb
  {\bf Lemma}\label{FS}\quad  
  If rules $R$ or $\omega$ take {\tt P} into ${\tt P}^*$ and $o\,{\tt 
  P}=\lambda<\varepsilon_0$
  then $o\,{\tt P}^*=\lambda_n$.
%   
%   
%   Assume ${\tt P=\langle^c\langle AT\rangle U}$ and 
%   $o\,{\tt P}=\lambda<\varepsilon_0$. We then have
%   \[
%  {\tt P}:x\ra_{\mbox{\FN{rule $R$ or $\omega$}}}{\tt P}^*:x\quad\mbox{implies}\quad
%  o\,{\tt P}^*=\lambda_n.
%  \]
  \ee
  \proof
  Assume that  {\tt P} is  in the form\n\ref{LIF}(\ref{NP-long}) with 
  \[o\,{\tt Q}=\gamma;\quad o\,{\tt R}_i=\beta_i;
  \quad o\,{\tt S}=\alpha.
%   \qquad\qquad\qquad\qquad(\mbox{with }o\,\epsilon=0).
  \]
 Case 1. ${\tt Q}\neq\epsilon$. We then have 
 ${\tt P}^*=\ang{\ang{\ldots\ang{\ang{Q}^{\it l\/}R_1}\ldots}R_c}\ {\tt S}$,
 and
   \[
    \lambda=\alpha+
  \omega^{\beta_c+\omega^{\ldots^{\omega^{\beta_1+\omega^{\gamma+1}}}}};\qquad
  \quad o\,{\tt P}^*=\alpha+
  \omega^{\beta_c+\omega^{\ldots^{\omega^{\beta_1+\omega^{\gamma}l}}}}=\lambda_n.
  \]
Case 2. ${\tt Q}=\epsilon$. We then have 
 ${\tt P}^*=\ang{\ang{\ldots\ang{ l\,R_1}\ldots}R_c}\ {\tt S}$, and
  \[
    \lambda=\alpha+
  \omega^{\beta_c+\omega^{\ldots^{\omega^{\beta_1+\omega}}}};\quad
  \quad o\,{\tt P}^*=\alpha+
  \omega^{\beta_c+\omega^{\ldots^{\omega^{\beta_1+l}}}}=\lambda_n.
  \]
%   \[
%     \lambda=\alpha+
%   \omega^{\beta_c+\omega^{\ldots^{\omega^{\beta_1+\omega}}}}\quad
%   \quad o\,{\tt Q}=
%   \omega^{\beta_c+\omega^{\ldots^{\omega^{\beta_1+l}}}}=\lambda_n.
%   \]
     \begin{nn}\label{thehierarchy}
      	    	\em 
      	    	{\bf The Hierarchy}\quad 
%       	    Let		$\db{{\tt P  }}:{\bf D}\mapsto{\bf D}$
%       	    	      	    		denote the  function computed by {\tt P}.
%       	    	      	    		,
%       	    	      	    		such that  $\db{{\tt P  }}(x)={\tt P  }(x)$.
% \section{hierarchy}       	  
      	    Define	 ($\db{{\tt P  }}$ is the function computed by {\tt P}) 
      	    \[
      	    \begin{array}{rcl}
      	    	{\cal C}&=&\{\db{{\tt P  }}\;|\;{\tt P  }\in 
      	    {\bf A}^*\ \mbox{for some {\bf A}}\};\\
      	    \qquad   
      	    {\cal C}_\alpha&=&\{\db{{\tt P}}\in{\cal C}\;|\;
      	    o\;{\tt P  }\leq\alpha\}.
      	    \end{array}      	    	    
      	    \] 
      	    {\bf Notation}\quad
       	 		 $\FN{TIMEF}(f(n))$ is the class of all functions 
 	$\varphi:{\bf D}\mapsto{\bf D}$
  	(for some {\bf D})  that are \tms-computed in deterministic time $O(f(|x|))$
   	for all $x\in{\bf D}$.
       	 \ee

\bb\label{WS}\quad Recall that  the Wainer Schwichtenberg majorization functions are
defined by
\[
F_1(n)=2n+1;\qquad
F_{\alpha+1}(n)=F_\alpha^{n+1}(n);\qquad F_\lambda(n)=F_{\lambda_{n}}(n).
\]
{\bf Claim}\quad We have ($c\geq 2$;  $l\geq 3$)
\ee 
\[\noind\!\!
\begin{array}{llllllllll}
{\rm (a)}& F_2(n)=2^ll-1;\qquad
{\rm (b)}&  F_2^c(n)\leq l_c;&\qquad&
{\rm (c)}& (l+2)_l\leq F_3(l).
\end{array}
\]
\proof (a) We show $F_1^c(n)=2^cl-1$. Induction on $c$. Step. 
$ F_1^{c+1}(n)=\IH 2(2^cl-1)+1$.

(b) One may believe this by a few computations for
$n=c=2$;  and by noting that $2^nn$ grows less than $2^{n+|n|+1}$ while
$n^n$ grows like $2^{n|n|}$ (for $n$  in binary). 

 (c) We first show by  induction on $c$
 \begin{equation}
 	 \label{F3}
 	 (l+2)_c\leq 2_c[l^2+lc].
 \end{equation}  
 Basis. For $l=3$, the assertion follows by computations;  for
 $l\geq 4$ by observing that $(l+2)^{l+2}$ grows like $2^{l|l|}$, obviously less than $2^{l^2}$.
Step.
We have 
\[\!\!\!\!
(l+2)_{c+1}\leq\IH (l+2)^{2_c[l^2+cl]}
\leq_{{\rm since\ }l+2\leq 2^l{\rm \ and\ }c\geq 1}
 2^{2_c[l^2+cl +l]}\leq 2_{c+1}[l^2+(c+1)l].
\]
The claim now follows because part (a) gives $F_2(l)\geq 2l^2$ and, therefore,
\[
F_3(l)=F_2^{l+1}(l)\geq F_2^l(2l^2)\geq_{{\rm part\ (a)\ }l\ 
{\rm times}} 2_l[2l^2]\geq_{{\rm by\ (\ref{F3})\ with\ 
}c=l} (l+2)_l. 
\]

     \section{Simulations}

\bb\label{ofTMs}
{\bf Simulation of TMs}\quad
We  restrict ourselves to  \TM  $M$ described by  arrays of size $q\times 
k^d$
whose elements $m(i,\vec{o})$
are $(d+1)$-ples  $(i^*,I_1,\ld,I_d)$,
with $1\leq i^*\leq q$ and $-1\leq I_h\leq k$ ($1\leq h\leq d$).
The  array says that $M$  has $q$ states and $d$ tapes, infinite to the right,
over an alphabet $\{S_1,\ld,S_k\}$.
If  $\vec{o}$ is scanned   in state $i$, $M$  enters  $i^*$, and
 moves left/right or writes  $S_l$ 
on (tape) $h$, according to  cases 
$-1,0,l$ of $I_h$.
Expression
 $D_{XT}=(i,l_1,o_1,r_1,\ld,l_d,o_d,r_d)$
says that, 
before step $T$ of the computation on the input $d$-ple $X$,
   string $l_ho_hr_h$ is  in  $h$,
 and  $o_h$ is scanned  in state $i$. 
Take as {\bf D} the  $(3d+2)$-ples of strings  over
$\Gamma=\{1,\ld,q,S_1,\ld,S_k\}$, and let $|x|$ be the max among the lengths 
of the strings of $x$. 
Define a {\em representation\/} of $D_{XT}$ in {\bf D}
by $x_{XT}=(i,l_1,o_1,r_1^R,\ld,l_d,o_d,r_d^R,1^{T+|X|})$,
where $r^R$ is $r$ read backward.
Take as {\bf A}  the  finite class of all
combinations of de/constructors on $\Gamma$ 
decided by the values of $q$ and  $\vec{o}$.
One of them is a program ${\tt nxt}^M$  taking
 any $x_{XT}$ into $x_{X,T+1}$.
Note that  the last component 
 of $x_{XT}$  ensures that we have 
 $|{\tt nxt}^M(x)|=|x|+1$.
\ee

        	    		\bb  
 	{\bf  Simulation by  TMs}\quad
Given {\bf A}, 	a \TM $M_{\bf A}$
 	 	 	with $t\geq 2$ 
 	 	 	tapes over  an alphabet ${\bf C}\supset{\bf A}$ 
 	 	 can be defined,	which simulates  all initial programs  in  1  step
 	 	(since their number is finite). 
Also, a \TM $M_{\bf A}^*$ with $t+3$ tapes can be defined,   that	 simulates any 
${\tt P}\in{\bf A}^*$
  in the way   outlined below.
It uses: tapes  	   $1,\ld,t$ \  for rule $A$ (via $M_{\bf 
 	 A}$),
 	 and  for keeping all parsing operations
 	 in linear time; 
 	    tapes $B=t+1$ and $S=t+2$ (resp.) for the initial programs  postponed
 	    (according to rule $P$), and
 	     for the current program;        	    		
       	    and 	 tape $t+3$  for  $l$, 
       	     and for a flag  $F$, which is set (cleared)
       	       	    if the program stored in $S$ is (not) safe.
%        	    
%        	    
%        	    and for a map  $F$ ($|F|\leq|{\tt P}|$) of the
%        	    repetitions depths, allowing to decide whether the last nested
%        	    repetition has been removed and, therefore, 
%        	    the program stored in $S$ is (now) safe.
       	     
 \begin{tabbing}
% KILL  KILL  KILL KILL  KILL  KILL KILL  KILL  KILL KILL  KILL  KILL
       	    		       \qquad	{\it while\/}  $S=U\,\up{\tt Q}$\qquad \=
       	    		 {\it if\/} ${\tt Q=T;1}$\qquad\qquad\qquad \={\it then push\/} 
       	    		 $(\up{\tt T})^N$ in $S$;\= end\kill%KILL
%  %%%%%%%%%%%%%%%%%%%%%%%%%%%%%%%%%%%%%%%%%%%%      	    		 
       	    		 \qquad$B:=\epsilon$; $S:={\tt P}$; \
       	    		 $l:=|x|+1$;\
                     assign $F$;\\*
       	    		 \qquad{\it while\/} $S\neq\epsilon$ {\it do\/}
       	    		 \>{\it if\/} $S=\langle^c\langle{\tt Q\,A}\rangle\,{\tt U}$\>{\it 
       	    		 then\/} $S:=\langle^c\langle{\tt Q}\rangle^l\,{\tt U}$; 
       	    		 update $F$\\*
       	    		\>{\it if\/} $S=\langle^c\langle{\tt A}\rangle\,{\tt U}$\>{\it 
       	    		 then\/} $S:=\langle^c{\tt A}^l\,{\tt U}$; 
       	    		 update $F$\\*
 \>{\it if\/} $S={\tt A\,Q}$ {\it and\/} $F=0$\>{\it 
       	    		 then\/} $S:=
       	    		 {\tt Q}$; apply $M_{\bf A}$\\* 
       	    		  \>{\it if\/} $S={\tt A\,Q}$ {\it and\/} $F=1$\>{\it 
       	    		 then\/} $S:=
       	    		 {\tt Q}$;  $B:=B\,{\tt A}$\quad{\it end-while\/}\\* 
       	    		 %
%  \>{\it if\/} $S=U\,{\tt \s 1}$\>{\it 
%        	    		 then\/} $S:=U{\tt q}$\>\ {\it end-while\/};\\*
       	    		\qquad apply $B$.
       	    	       	    		        	    		  \end{tabbing}
       	    	       	    		        	    		  \ee
%        	    		        	    		  \nopagebreak
%          	    		        	    		 \LINE 

% \section{runtime}	    

% 	  \newpage \newpage \newpage \newpage \newpage \newpage \newpage       	    	
    
\bb  
\label{runtime}  {\bf Functions size and simulation runtime} \ (1) \ For all $\alpha$ define
the  functions: 

(a)\quad 
    $R_\alpha(n)$, returning 
the time for the simulation   of  any {\tt P}  on $x$
($o\,{\tt P}=\alpha$);

(b)\quad 
 $\tilde{h}_\alpha(n)$ such that $o\,{\tt P}=\alpha$\quad
  implies\quad ${\tt P}(x)=_c\tilde{h}_\alpha(n)$
  (by induction on $\alpha$ one sees that $\tilde{h}_\alpha$
 is well defined, in the sense that
 $ o\,{\tt P}=o\,{\tt Q}=\alpha\quad\mbox{implies}\quad|{\tt P}(x)|=|{\tt 
 Q}(x)|$).

(2)\quad We claim that we have ($0\leq\beta$; $\beta+\lambda<\varepsilon_0$) 
      \[  \noind\!\!
     \begin{array}{lrllllllllll}
    1\quad&R_{\beta+m}(n)&=&R_\beta(n)+m
    \\
 2& R_{\beta+\lambda}(n)&\leq& 
  	R_{\beta+\lambda_n}(n)+l(|\lambda|-1)
  	\ \leq\ R_{\beta+\lambda_n}(l)\qquad
%   	&(|\gamma|=1
%   	\mbox{ when rule $\omega$ is applied})
%    	\\&&\leq&R_{\beta+\omega^\gamma l}(l)&\mbox{if}&\gamma\geq\omega
   \\
 3&  R_{\beta+\omega}(n)&\leq&R_\beta(n+l)+l&\mbox{if}&\beta\geq\omega^\omega
    \\
    4&  R_{\beta+\omega}(n)&\leq&R_{\beta}(n)+2l&\mbox{if}&\beta<\omega^\omega.
    \end{array}	
%      	\label{eq-runtime}
     \]
   We discuss line  2 only because the others are rather obvious.
   By  working simultaneously  on all tapes, and by linear speed-up 
    techniques,  $M^*_{\bf A}$ can update $F$ and produce $l$ copies of a program shorter than 
    $|\lambda|$   
     in  $l(|\lambda|-1)$ steps.
      By Lemma\n\ref{FS} we may express this in terms of $\lambda$
     and $\lambda_n$. 
%       Line 4. By a linear speed-up by a factor 3 we can simulate in $l$ 
%       steps rules $\omega$ (once) and $A$ ($l$ times). 

%      We have the  second inequality because
%       $R_{\beta+\lambda}(n)$ grows 
%      faster than $l^{|\lambda|}$.
%   (4)\quad Note that, if {\tt A} is the only initial program occurring in {\tt P},
%   and $o\,{\tt P}=\alpha$ then 
% \[
%    {\tt P}(x)={\tt A}^{\tilde{h}_\alpha(n)}(x).
%   \]
          \ee

 % \section{composition}
\bb\label{composition}{\bf Composition lemma}\quad
%  Assume 
% $\beta+\gamma<\varepsilon_0$. 
We have
\[\noind\!\!
\begin{array}{lrllllllllllll}
	\mbox{(1A)}& 
{\tilde h}_{\beta+\gamma}(n)&=&{\tilde h}_\beta(n+{\tilde 
h}_\gamma(n))&\mbox{if}&\beta&\geq&\omega^\omega;\\
\mbox{(1B)}&{\tilde h}_{\beta+\gamma}(n)&=&{\tilde h}_\beta(n)+{\tilde 
h}_\gamma(n)&\mbox{if}&\beta&<&\omega^\omega;\\

\mbox{(2A)}&R_{\beta+\gamma}(n)&\leq& 
R_\beta(n+\tilde{h}_\gamma(n))+R_\gamma(n)
\leq  R_\beta(R_\gamma(n)+1)
& \mbox{if}&\beta&\geq&\omega^\omega;
\\ 
\mbox{(2B)}&R_{\beta+\gamma}(n)&\leq& 
R_{\beta}(n)+R_\gamma(n)&\mbox{if}&\beta&<&\omega^\omega.
% \\
% \mbox{(3)}&\tilde{h}_{m+\beta}(n)&=&\tilde{h}_\beta(n)+m.
\end{array}
\]
  \ee
\proof
Immediately by the discussion at the end of\n\ref{comment}
(with $o\,{\tt S}=\gamma$ and $o\,{\tt T}=\beta$).

The second inequality of (2A) follows from our estimate $R_\alpha(n)\geq 2^n$ 
at $\alpha\geq\omega^\omega$.

\bb{\bf Lemma}\label{sandwich}\quad
  $F_\beta(n-d)\leq\tilde{h}_{\omega^\alpha}(n)\leq R_{\omega^\alpha}(n)
  \leq F_\beta(n+e+1)\quad $   implies
 \[
 F_{\beta+1}(n-d)\leq\tilde{h}_{\omega^{\alpha+1}}(n)\leq R_{\omega^{\alpha+1}}(n)
  \leq F_{\beta+1}(n+1)
  \quad (d,e\leq 1;\,\beta\geq 
  2;\, \alpha\geq\omega).
 \]
  \ee
 \proof
 First half. Immediately  by  the strict monotony of all 
 $F_\beta$, since $n-d\geq 1$.
 
 Second half. We first show $R_{\omega^{\alpha }c}(n)\leq F_\beta^c(n+c+e+1)$.
Induction on $c$. Step. 
\[\noind\!\!\!
\begin{array}{lllr}
	R_{\omega^{\alpha }(c+1)}(n)&\leq&  
 R_{\omega^{\alpha }}( R_{\omega^{\alpha }c}(n)+1)&\mbox{L. 
 \ref{composition}(2A)}\\
  &\leq& R_{\omega^\alpha}(F_\beta^c(n+c+e+1)+1)&\mbox{I.H.}\\
   &\leq& F_\beta(F_\beta^c(n+c+e+1)+e+1)\leq F_\beta^{c+1}(n+c+e+1)&
   \mbox{hypotheses}
%    \\
%     &&\leq& &\mbox{hyp. $\beta\geq 2$}\\
\end{array}
\]
Our assertion now follows because we have
\[
R_{\omega^{\alpha+1}}(n)\leq_{{\rm \ref{runtime}(2)2}} R_{\omega^{\alpha}l}(l)\leq
F_{\beta}^l(n+l+e+1)\leq F_{\beta}^{l+1}(l)\leq F_{\beta+1}(l). 
\]
%  , and, therefore  
% \[
% 2^n\leq F_\beta(n-d)\leq F_\beta(n-d+F_\beta(n-d)).
% \]
% 

% \section{poly}
\bb\label{poltime}
{\bf Lemma} \ ({\bf Polytime})  For all $c\geq 1$ we have
\[
\begin{array}{lrlllllllll}
&l^c&=&\tilde{h}_{\omega^c}(n)&\leq& R_{\omega^c}(n)&\leq& 
l^c+lc^2.
\end{array}
\]  
 \ee
\proof We show by induction on $c$ that we have 
\[
\tilde{h}_\beta(n)+l^c=\tilde{h}_{\beta+\omega^c}(n)\leq
R_{\beta+\omega^c}(n)\leq R_\beta(n)+l^c+c^2l
\qquad(0\leq\beta<\omega^\omega).
\]
 Basis. One half by\n\ref{composition}(1B), since rules
  $\omega$ and $P$  give 
        $\tilde{h}_{\beta+\omega}(n)=\tilde{h}_{l+\beta}(n)$.
  The other half  by\n\ref{runtime}(2)4.
            Step.    
  \[\noind\!\!
      \begin{array}{lllrlllllllllllll}
\tilde{h}_{\beta+\omega^{c+1}}(n)&=&
\tilde{h}_{\beta+\omega^{c}l}(n)&\mbox{rule $R$}\\
&=&\tilde{h}_{\beta+\omega^cn}(n)+l^c\quad&\mbox{I.H. with 
$\beta+\omega^cn$ as $\beta$}\\         
        &=&\tilde{h}_\beta(n)+ll^c&\!\!\!\!\mbox{I.H. $n$ times more, with 
$\beta+\omega^ci$ as $\beta$}.\\
R_{\beta+\omega^{c+1}}(n)&	\leq& R_{\beta+\omega^{c}l}(n)+(c+1)l&
	\mbox{\ref{runtime}(2)2}
	\\
	 &	\leq& R_\beta(n)+ll^c+c^2l+(c+1)l&
	 \mbox{I.H. $l$ times}\\
&	\leq& R_\beta(n)+l^{c+1}+(c+1)^2l.
\end{array}
  \]
%    \newpage
%   \newpage\newpage\newpage\newpage\newpage\newpage\newpage\newpage
\bb\label{MAJ-EXP}{\bf Functions Size at the Elementary Level}\quad
Define $h_c(n)=n+\tilde{h}_{\omega^\omega c}(n)$. 
 
 Lemma\n\ref{poltime} with 
$h_c(n)$ as both $c$ and $l$ gives
% (since $\tilde{h}_{\omega^\omega(c+1)}(n)=\tilde{h}_{\omega^\omega c}(n)_1$)
\begin{equation}
	\label{h-c}
	h_1(n)=n+l^l;\qquad h_{c+1}(n)= h_{c}(n)^{h_{c}(n)}+h_{c}(n)
\end{equation}
 To majorize $h_c$ define a sequence of functions $H_c$ by
       \begin{equation}\label{defH}
 \begin{array}{rllllllll}
  {H}_0(l)=l;&  
  {H}_1(l)=l^{1+l};&{H}_2(l)=l^{l^{2+l}};\quad
    {H}_{c+3}(l)\ =\  l^{l^{l^{1+l^{\ldots^{1+l^{2+l}}}}}}
		     \end{array}
		            \end{equation} 
	($l$ thrice,  $1+l$ for
     $c-3\geq 0$  times, and $2+l$ once). 	    Thus for $c\geq 3$ we have
		    		    \begin{equation}
	    H_c(l)=l_3[q]\quad\mbox{for some }q;\quad
			    H_{c+1}(l)=l_3[1+l^q]\quad\mbox{for the same }q.		
			    \label{H34}
			       		    \end{equation} 
					     For example, $H_3(l)=l_3[l+2]$ and $H_4(l)=l_3[1+l^{2+l}]$.
					     
    {\bf Claim}\quad $h_c(n)\leq H_c(l)$.
    
    \proof Induction on $c$. Basis.
%     We have three cases. 
     $c=1$. By (\ref{h-c}), since $n+l^l\leq 
    l^{1+l}$.
    To discuss   next two cases of the basis, define $L=l^{1+l}+l^l+nl+n$ and note that    
        \begin{equation}
	L+3+l\leq l^{1+l}+l^l+2l^2\leq l^{1+l}+l^l+l^3\leq 2l^{1+l}\leq l^{2+l}.       
	\label{L}
	 \end{equation}
    For    $c=2$, we have by (\ref{h-c}), twice with $c=1$ and $c=2$, and (\ref{L}),
    since $n+l^l+1\leq l^{1+l}$ 
    \[
    h_2(n)\leq
    (n+l^l)^{(n+l^l)}+(n+l^l)\leq (l^{1+l})^{(n+l^l)}\leq l^L\leq
    l^{l^{2+l}}.
    \]
    For $c=3$, we have by (\ref{h-c}), case $c=2$, and by (\ref{L}) 
    \[
   h_3(n)\leq (l^{l^{2+l}})^{l^L}+l^{l^{2+l}}\leq (l^{l^{3+l}})^{l^L}
   \leq l^{l^{3+l+L}}
    \leq l^{l^{l^{2+l}}}.
    \]
    Step. 
    By (\ref{H34}) we have $H_c(l)=l_3[q]$ for some $q$. Hence
    \[
    h_{c+1}(n)\leq\IH (H_c(l))_1+H_c(l)%\leq l_3[q+1]^{l_3[q]}
     \leq l_3[1+l^q]
    =_{{\tt by\ (\ref{H34})}} H_{c+1}(l)
    \]
    where  we have  last inequality because
    \[
    (l_3[q])_1+l_3[q]=(l^{l^q})^{l^{l^q}}+l^{l^q}\leq l^{l^q l^{l^q}+1}
	    \leq l^{l^{l^{1+q}}}=l_3[1+q].
    \]
    \ee

% \section{supexp}

\bb\label{supexp}
{\bf Lemma} \ ({\bf  Superexptime})  For all $c\geq 1$ we have
\[
\begin{array}{lrlllllllll}
&l_c&\leq& {h}_{\omega^\omega c}(n) &
  \leq&R_{\omega^\omega c}(n)&\leq& 
      (l+2)_c.
\end{array}
\]  
 \ee
\proof Both halves by induction on $c$.
 First half.   Basis.  Example\n\ref{Ex1}.4 gives
  $\tilde{h}_{\omega^\omega}(n)= l^l$.
    Step. By arguments like in proof of
 Lemma\n\ref{sandwich} (with $(n+1)_1$ instead of  $F_\beta(n)$).
     
   Second half. Basis. By Lemma\n\ref{poltime} and \ref{runtime}(2)2 since we  have 
   $l^l+l^3+2l\leq l^{2+l}$.
   Step. 
   \[\noind\!\!
   \begin{array}{lllr}
   	R_{\omega^\omega (c+1)}(n)&\leq& 
   	R_{\omega^\omega}(H_c(n))+R_{\omega^\omega c}(n)&
   	\mbox{\ref{composition}(2A), Claim\n\ref{MAJ-EXP}}\\
   	&\leq&(H_c(n))_1+H_c(n)^3+2l+R_{\omega^\omega c}(n)&\!\!\mbox{like under 
   	the basis}\\
   	&\leq&(H_c(n))_1+H_c(n)^3+2l+(l+2)_c&\mbox{I.H.}\\
   	&\leq&2^{(l+2)_c}+(l+2)_c^3+2l+(l+2)_c\leq(l+2)_{c+1}.
   \end{array}
   \]
%    
%    
%    
%    
%    
%    
%     We  use the claim\n\ref{MAJ-EXP} to  show
%  \begin{equation} 
%  \begin{array}{l}
%  	R_{\omega^ \omega c}(n)\leq \sum_{1\leq i\leq 
% 	 	c}({H}_i(l)+2{H}_{i-1}(l)^3). \end{array}
%   	\label{majR}
%  \end{equation}
% %  Because of the majorization applied later to the estimate (\ref{majR}) we may 
% %  fairly restrict ourselves to the step only. 
%  Induction on 
%  $c$. We have
%  \[\noind\!\!
%  \begin{array}{rllr}
%  R_{\omega^ 
%  	\omega}(n)&\leq&R_{\omega^l}(n)+2l\leq l^l+l^3+2l\leq H_1(l)
%  	+2H_0(l)^3\quad& 
% %  	\!\!\!\!\!\!\!
%  	\mbox{\ref{runtime}(2)2, L.\n\ref{poltime}}
%   	\\
% %  	&\leq&l^{l+1}+l&\mbox{I.H.}\\
%  	R_{\omega^ 
%  	\omega(c+1)}(n)&\leq&R_{\omega^\omega}(n+{H}_c(l))%+{H}_c(l)^3
%  	+R_{\omega^\omega c}(n)&
% %  	\!\!\!\!\!\!\!
%  	\mbox{L.\ref{composition}(2A), Cl. \ref{MAJ-EXP}}
%  	\\
%  	&\leq&{H}_{c+1}(l)+{H}_c(l)^3+
%  	 \sum_{i\leq 
%  	c}({H}_i(l)+2{H}_{i-1}(l)^3)&\mbox{I.H.    and L.\n\ref{poltime}}
%  \end{array}
%  \]
%  From ${H}_c(l)^3\leq{H}_{c+1}(l)$
%  we obtain $\sum_{i\leq 
%  	c}({H}_i(l)+2{H}_{i-1}(l)^3)\leq 3c{H}_c(l)$.
%  	The result follows because for $l\geq 3$ we obviously have
%  	$3c{H}_c(l)\leq(l+2)_c$.

% \section{gregor}

    \bb\label{gregor}{\bf Lemma} \ ({\bf Gregorczyk  classes}) 
     For  all $c\geq 1$  we have
         \[
     \begin{array}{lrlllllllllllllllll}
& F_{c+2}(n)&\leq& \tilde{h}_{\omega^{\omega+c}}(n)&\leq&
   R_{\omega^{\omega+c}}(n)&\leq& F_{c+2}(n+1).
   \end{array}
 \] 
\ee       
     \proof 
 Induction on $c$. Basis. Rule $R$ (with  ${\tt U}=\epsilon$) takes 
      $\ang{1\ang{1}}$ into $\ang{\ang{1}}^l$. Hence 
      \[
      \begin{array}{rl}
     \tilde{h}_{\omega^{\omega+1}}(n)&= \tilde{h}_{\omega^\omega l}(n)
     \geq_{{\rm L.\ \ref{supexp}}}l_l
     \geq F^l_2(n)_{{\rm L.\n\ref{sandwich}\   by \
      \ref{WS}(b)}}=F_3(n)\\
   	R_{\omega^{\omega+1}}(n)&\leq_{{\rm \ref{runtime}(2)2%,\ 1st line
   	}}
   	R_{\omega^{\omega}l}(n)+3l\leq_{{\rm L.\n\ref{supexp}}}
   	   	 (l+2)_l\leq_{{\rm \ref{WS}(c)}} F_3(l).
   \end{array}
   \]
where, in the second last inequality,  we can ignore term $3l$ because
 neglectable with respect to the majorization at the end  of last proof.   

    Step.
Rule $R$ takes   $\ang{1c\ang{1}}$ into  $\ang{c\ang{1}}^l$,
with $o\,\ang{c\ang{1}}^l=\omega^{\omega +c}l$. 
The result follows by Lemma\n\ref{sandwich} and I.H., with $c+2$ as $\beta$, 
 $\omega+c$ as $\alpha$, and $e=1$¥.

 \bb\label{2-nested}{\bf Lemma} \ ({\bf  2-nested recursion classes}) 
     For  all $c\geq 1$, and $d\geq 0$ we have
         \[ F_{\omega c+d}(n)\leq \tilde{h}_{\omega^{\omega (c+1)+d}}(n)\leq
   R_{\omega^{\omega (c+1)+d}}(n)\leq F_{\omega c+d}(n+2)  
    \] 
\ee 
\proof Induction on $\omega(c+1)+d\geq\omega 2$.
 Basis.
$d=0$ and $c= 1$. 
\[%\noind\!\!
\begin{array}{rllr}
\tilde{h}_{\omega^{\omega 2}}(n)&=
\tilde{h}_{\omega^{\omega+l}}(n)\geq_{{\rm L.\ \ref{gregor}}}
  F_{l+2}(n)\geq F_l(n)=F_{\omega}(n)
\\
	R_{\omega^{\omega 2}}(n)&\leq_{{\rm \ref{runtime}(2)2}}
	R_{\omega^{\omega+l}}(l)
	\leq _{{\rm L.\ \ref{gregor}}} F_{l+2}(n+2)=F_{\omega }(n+2).
\end{array}
\]
Step. Case 1.
$d=0$ and $c> 1$. 
% By I.H. and \ref{runtime}(2)2 (for the second half) 
We have
\[\noind\!\!
\begin{array}{l}
\tilde{h}_{\omega^{\omega(c+1)}}(n)=
\tilde{h}_{\omega^{\omega c+l}}(n)
\geq\IH F_{\omega(c-1)+l}(n)=F_{\omega c}(n)
\\
	R_{\omega^{\omega (c+1)}}(n)\leq_{\ref{runtime}(2)2}¥
	R_{\omega^{\omega c+l}}(l)
	\leq\IH F_{\omega(c-1)+l}(l+2)
% \leq F_{\omega(c-1)+l+1}(n+3)\\
\leq
	F_{\omega(c-1)+l+2}(l+1)  =  
	F_{\omega c}(l+1)
\end{array}
\]
Case 2. $d\geq 1$.
Rule $R$ takes   $\ang{1d\ang{1}^{\it c\/}}$ into  $\ang{d\ang{1}^{\it c\/}}^l$,
with $o\,\ang{d\ang{1}^{\it c\/}}^l=\omega^{\omega c+d}l$. 
The result follows by Lemma\n\ref{sandwich} with $\omega(c-1)+d$ as $\beta$, 
 $\omega c+d$ as $\alpha$, and $e=2$.

% \section{Large}

    \bb\label{large}{\bf Lemma} \ ({\bf Large classes}) 
     $\omega^2\leq\alpha
   <\varepsilon_0$ implies
         \[
    F_{{\alpha}}(n-1)\leq \tilde{h}_{\omega^{\alpha}}(n)
   \leq
   R_{\omega^\alpha}(n)\leq F_{{\alpha}}(n+1)¥.	
     \] 
\ee       
     \proof 
   Induction on $\alpha$. 
     Basis. $\alpha=\omega^2$. We have
\[
\begin{array}{rllr}
	\tilde{h}_{\omega^{\omega^2}}(n)&=&
\tilde{h}_{\omega^{\omega  l}}(n)\geq
 F_{\omega n}(n)&{\mbox{Lemma \ref{2-nested}}}\\
&\geq& F_{\omega n}(n-1)=F_{\omega^2}(n-1).\\
% %   
R_{\omega^{\omega^2}}(n)&\leq&
R_{\omega^{\omega l}}(l)&\mbox{\ref{runtime}(2)2}
\\&\leq& F_{\omega n}(l+2)&\mbox{Lemma \ref{2-nested}}
\\&\leq& F_{\omega n+1}(l)\leq F_{\omega (l+1)}(l)=  F_{\omega^2}(l).\qquad\qquad\qquad\qquad
\end{array}
\] 
Step. Case 1. $\alpha$ is a limit $\lambda$.
% Either rule $R$ or rule $\omega$ applies.
%   By   Lemma\n\ref{FS} 
% the ordinal of the program they  produce is $\lambda_n$. So we can apply the I.H.
\[
\begin{array}{rllr}
	\tilde{h}_{\omega^\lambda}(n)&=&
\tilde{h}_{\omega^{\lambda_{n}}}(n)\geq\IH F_{\lambda_{n}}(n-1)\geq
 F_\lambda(n-1)\\
R_{\omega^{\lambda}}(n)&\leq&
R_{\omega^{\lambda_{n}}}(l)_{\ref{runtime}(2)2}
\leq\IH F_{\lambda_{l}}(l+1)
\leq  F_{\lambda_{l+1}}(l)= F_{\lambda}(l).
\end{array}
\] 
   Case 2. $\alpha=\gamma+1$.
   The result follows by Lemma\n\ref{sandwich} with $\gamma$ as both $\beta$ and
 $\alpha$.
% 
%    
%     We have
%    \[\noind\!\!
%    \begin{array}{rclr}
%    	\tilde{h}_{\omega^{\beta+1}}(l)&=&\tilde{h}_{\omega^\beta(l+1)}(l)
%    	\geq\tilde{h}_{\omega^\beta}^{l+1}(l)&\mbox{Lemma \ref{composition}(3)}\\
%    	&\geq&F^{l+1}_\beta(n)\geq F^l_\beta(n)=F_{\beta+1}(n)&\mbox{I.H.}\\
% %    	
% R_{\omega^{\beta+1}}(l)&\leq&R_{\omega^\beta(l+1)}(l+1)
% &\mbox{\ref{runtime}(2)2}\\
%    	&\leq&F_{\beta}^{l+1}(2l+2)&
%    	\!\!\!\!\!\!\!\!\!\!\!\!\!\!\!\!\!\!\!\!\!\!\!\!\!\!
%    	\mbox{L.\ref{composition}(3) with $F_\beta$ 
%    	as $f$ by I.H., $c=l+1$,   $d=1$}\\
% %    	&\leq&F^{l+1}_\beta(3l+3)&\!\!\!\!\!\mbox{L. \ref{composition}(3) with $F_\beta$ 
% %    	as $f$ by I.H.}\\
%    	&\leq&F^{l+2}_\beta(l+1)=F_{\beta+1}(l+1).
%    \end{array}   
%    \]
  
%  	\section
\bb{\bf Proof of the Theorem}\label{proof-th}\quad
Inclusions of the form $\FN{TIMEF}(f_\beta(n-1))\subseteq{\cal C}_\alpha$.
% ($e=1$ for $\alpha\geq\omega^{\omega^2}$).
For each function $\varphi\in\FN{DTF}(f_\beta(n-1))$
% be given.
%  By linear speed-up results,
there is a \TM $M$ in the form 
of\n\S\ref{ofTMs} that computes $\varphi$ within $f_\beta(n-1)$ steps.
Let ${\tt nxt}_\alpha^M$ denote the program whose ordinal is $\alpha$
and in which no other initial program but ${\tt nxt}^M$ occurs.
% By \ref{runtime}(4), ${\tt nxt}_\alpha^M(x)$ iterates ${\tt nxt}_M$ on $x$
% for $\tilde{h}_\alpha(n)$ times.
 Our assertions follow by 
Lemmas\n\ref{poltime}, \ref{supexp},
\ref{gregor}, \ref{2-nested} and \ref{large} 
since: they state  $f_\beta(n-1)\leq\tilde{h}_\alpha(n)$;
 and since if {\tt A} is the only initial program occurring in {\tt P},
we obviously have 
${\tt P}(x)={\tt A}^{\tilde{h}_{o\,{\rm P}}(n)}(x)$.
%   \]

Other set of inclusions. The same lemmas show that
the time complexity of an interpreter for  language ${\bf A}^*$
respects the asserted  upper bounds.
\ee 
%  \vspace{-4mm}


\begin{thebibliography}{}
%  \vspace{-2mm}
	
{\footnotesize 

% \vspace{-2mm}
\bibitem[BC]{BC} S.J.Bellantoni and S.Cook. {\it  A new recursion-theoretic
characterization of the poly-time functions.\/} Computational Complexity
2(1992)97-110.
 
%$$ 
% \vspace{-2mm}
  \bibitem[CZG]{CZG}	  
 	 S.	Caporaso,  M. Zito,	N. Galesi. 
 	 {\em A	predicative	and	decidable  characterization	of the polynomial classes 
 	 of languages.\/}
 	   Theoretical Comp.	Sc. 250(2001)83-99
 %	% 
  
%   \vspace{-2mm}
   \bibitem[C]{C} S. Caporaso. {\it A decidable characterization of the 
   classes between lintime and exptime.\/} Inf. Proc. Letters 97(2006)36-40
   
%     \vspace{-2mm}\bibitem[C2]{C2} S. Caporaso.
%      Papers on ICC. On the web at
%      http:// www.di.uniba.it/intint/caporaso.html

%  \bibitem[B96]{BN} S.J.Bellantoni, {\it  Ranking primitive 
%  recursion: the low Grzegorczyck classes revisited\/}.
%  DIMACS Tech. Reports 96-50(1996).
% % 
% % 
% % 
% % 
%  \bibitem[C96]
%  {Cap}
%  S. Caporaso, {\em Safe TMs, Grzegorczyk classes and polytime.\/}
%  Intern. J. 	Found. Comp. Sc. 7.3(1996)241-252.

 
% 	   \bibitem[C92]{cichon}	E.A. Cichon, 
%    	   {\em Terminating proofs and complexity characterizations.\/}
%    	   In P. Aczel et al. (eds) Proof Theory. Cambridge U. Press
%    	   (1992)173-93.

% \vspace{-2mm}
 \bibitem[FW]{FW} M. Fairtlough and S.S. Wainer.
  {\em Hierarchies of provably 
 recursive functions.\/} In S. Buss (ed.) Handbook of Proof Theory. 
 Elsevier,  1998.		
 %
 	
  
%      \bibitem[L94]{Lpoly} D. Leivant, {\em A foundational delineation of 
%     computational feasibility.\/} Information and Computation 
%     110(1994)391-420.
% 
  
%   \vspace{-2mm}
   \bibitem[L]{LeivantT}
    D. Leivant. {\em Intrinsic Theories and Computational Complexity\/}, in D. Leivant
     (ed.), Logic and Computational Complexity.
    Springer, LNCS 960(1995)177-194.
    
     \vspace{-2mm}
   \bibitem[OW]{OW} G.E. Ostrin and S.S. Wainer. {\em Proof theoretic 
 complexity\/}. In H. Schwichtenberg and E. Steinbrugger (eds.)
 Proofs and System Reliability. Kluyver, 2002. 	
  
 

%% 
 % 

%  \bibitem[R]{Rose}
%  H.E. Rose.  {\em Subrecursion: Functions and Hierarchies\/}.
%  Oxford University Press,  	1984.

  \vspace{-2mm}
   \bibitem[W]{W} S.S. Wainer. {\em Ordinal recursion
   and a refinement of the extended
   Grzegorczyk hierarchy.\/} J. Symb. L. 37.2(1972)281-292.}
\end{thebibliography}
	 	\end{document}